\documentclass[a4paper,11pt]{article}
\usepackage{pos}
\usepackage{hyperref}

\usepackage{tikz}
\usepackage{color}
\usepackage{xcolor}

\title{Indirect Searches for Secluded Dark Matter}
 \ShortTitle{Indirect Searches for Secluded Dark Matter}

\author*[a]{C. Siqueira}
\author[a]{Guilherme N. Fortes}
\author[b,c,d]{Farinaldo S. Queiroz,}
\author[a]{Aion Viana}

\affiliation[a]{Instituto de F\'isica de S\~ao Carlos, Universidade de S\~ao Paulo, Av. Trabalhador S\~ao-carlense 400, S\~ao Carlos-SP, 13566-590, Brasil.}
\affiliation[b]{International Institute of Physics, Federal University of Rio Grande do Norte, Campus Universitário, Lagoa Nova, Natal-RN 59078-970, Brazil}
\affiliation[c]{Departamento de F\'isica, Universidade Federal do Rio Grande do Norte, 59078-970, Natal,
RN, Brasil}
\affiliation[d]{Millennium Institute for SubAtomic Physics at the High-energy frontIeR, SAPHIR, Chile}

\emailAdd{csiqueira@ifsc.usp.br}

\abstract{Dark matter is one of the most important open problems in particle physics and cosmology. Weakly interacting massive particles (WIMPs) appear as an appealing solution, providing the right relic density with a cross-section at the electroweak scale, however, no WIMP signals were observed until now. Secluded models are good alternatives to the standard ones. In this case, instead of a direct annihilation into the standard model (SM) particles, the dark matter annihilates into mediators which subsequently decay into SM particles. In this way, secluded models may avoid the stringent limits from direct searches, and, at the same time, be probed by indirect detection experiments. Motivated by the appearance of secluded dark matter in several model building endeavors, in this talk, we will present the sensitivity of several gamma-ray instruments (current and prospects), including Fermi-LAT, H.E.S.S., CTA, and SWGO, to secluded dark matter annihilations in the inner galactic halo, and in the dwarf spheroidal galaxies, covering a wide range dark matter masses, from tens of GeV to hundreds of TeV.}

\FullConference{37$^{\rm{th}}$ International Cosmic Ray Conference (ICRC 2021)\\
		July 12th -- 23rd, 2021\\
		Online -- Berlin, Germany}

\begin{document}
\maketitle

\section{Introduction}

Nowadays, we have several evidences for the dark matter (DM) existence, all of them points towards a particle which needs to be neutral, weakly interacting, massive, cold or warm, and sufficiently abundant
\cite{1991Natur.352..769P}. One of the main candidates and by far, one of the most studied in the literature, are the weakly interacting massive particle, the WIMPs. These candidates, with a mass between a few GeV to a few TeV, provide the right relic abundance in accordance with Planck satellite \cite{Aghanim:2018eyx} with an annihilation cross-section at the weak scale \cite{Arcadi:2017kky}. An interesting aspect of these WIMP particles is that they can be probed by indirect detection experiments. However, strong limits have been put over these candidates, including that from collider and direct searches \cite{Arcadi:2017kky}. 

Usually, interactions between DM and standard model particles are mediated by new forces. And the same coupling which gives the right relic abundance also mediate nucleon-WIMP scattering interactions. So, with the most recent results from XENON1T \cite{XENON:2018voc}, it is hard to provide the right relic abundance and escape from these stringent bounds, frequently, only the peak of the resonance region can be free from the current limits \cite{Arcadi:2017kky}. Therefore, the constrained WIMPs leads the scientific community to look for alternative scenarios, which include non-standard cosmologies \cite{Arias:2019uol}, new dark matter production mechanisms \cite{Bernal:2017kxu}, or even both \cite{Cosme:2020mck}, and so on. But there is also an interesting possibility to disconnect these observables, if the DM mass is larger than the mediator's, the DM can annihilate into mediators which belongs to the dark sector, and this channel may be more relevant, in this case, the DM is called secluded \cite{Pospelov:2007mp}. An interesting aspect from these scenarios is that they can escape from the current stringent bounds, and it also provides interesting indirect signatures \cite{Pospelov:2008jd,Profumo:2017obk,Leane:2017vag,Leane:2021tjj}. 

Given the motivation for secluded models, in this talk, we will focus on the indirect gamma-ray searches for secluded models into the galactic halo and into dwarf spheroidal galaxies, to explore what are the main constraints from indirect gamma-ray searches on these scenarios. We will use data from the current Fermi-LAT \cite{Ackermann:2015zua} and H.E.S.S. \cite{Abdallah:2016ygi} experiments, and also the prospects for the Southern telescopes CTA \cite{Acharyya:2020sbj} and SWGO \cite{Schoorlemmer2019}. This proceeding is structured as follows: in Section~\ref{sec:model}, we will discuss the secluded setup, in Section~\ref{sec:flux}, we will explore how to compute gamma-ray flux for DM particles in a secluded sector. In Section~\ref{sec:exp}, we describe the experiments used in this work, in Section~\ref{sec:res}, we present our results and in the last section we draw our conclusions.

\section{Secluded Models}
\label{sec:model}

Secluded scenarios were overly studied in the literature in context of the galactic center excess \cite{Boehm:2014bia,Yang:2020vxl}. Another interesting indirect signature is provided by the search for DM signals at planets and stars \cite{Leane:2017vag,Leane:2021tjj}, in these cases, when the DM annihilates directly into SM particles the only final state we can observe from indirect searches is an attenuated neutrino signal, because the other SM channels are trapped by the environment inside the stars/planets. However, if the DM annihilates into metastable mediators (with an enough decay rate), they can easily escape from the surface's planet/star, and its subsequently decay into SM particles leaves interesting indirect signatures \cite{Pospelov:2008jd}, including into gamma rays.

\begin{figure}[ht!]
    \centering
    \includegraphics[width=0.48\textwidth]{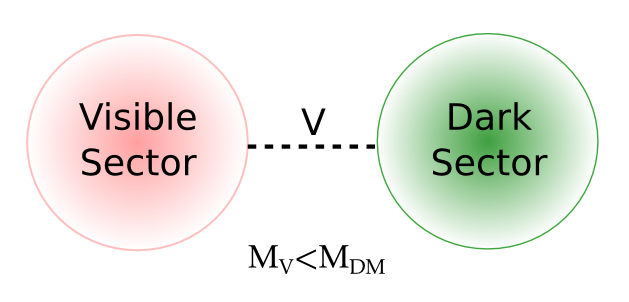}
    \includegraphics[width=0.30\textwidth]{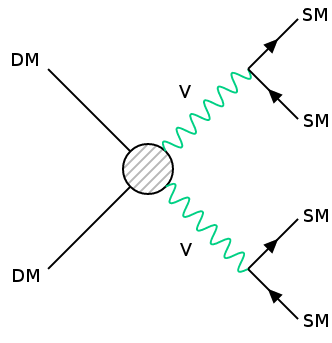}
    \caption{Schematic scenario for secluded models (right), Feynman diagram representing secluded interactions. }
    \label{fig:diag}
\end{figure}

As already explained, secluded models appears when instead of annihilate directly into SM particles, it annihilates into particles that belongs to the dark sector, see Fig.~\ref{fig:diag} (left). These mediators can be a fermion, a scalar or a vector \cite{Elor:2015tva}, and it can subsequently decay into SM particles. So, in the context of indirect searches, instead of a $2 \to 2$ annihilation, we have a $2 \to 4$ one, as can be seen in the right panel of Fig.~\ref{fig:diag}. We emphasize that there are many secluded DM models in the literature, which includes dark photons \cite{Pospelov:2007mp,Batell:2009zp,Dutra:2018gmv}, heavy vector bosons \cite{Dedes:2009bk,Fortes:2015qka}, and also scalar fields \cite{Yaguna:2019cvp,Belanger:2020hyh}.
After this brief description of the models, we will follow to the computation of the gamma-ray flux. 

\section{Dark Matter gamma-ray flux}
\label{sec:flux}

The gamma-ray flux expected by a DM annihilation is given by

\begin{equation}
 \frac{\phi_\gamma}{dE}= {\color{green}\overbrace{\color{black} \frac{\langle \sigma v \rangle }{8\pi m_{DM}^2} 
            \frac{dN_\gamma}{dE}}^{\mathrm{Particle\,\, Physics}}}
             {\color{purple} \overbrace{\color{black} \int ds \int d\Omega \, 
            \rho_{DM}^2}^{\mathrm{J-Factor}}}  \nonumber
\end{equation}
where we have a part which depends on the particle physics, that comprises the DM mass ($m_{DM}$), the velocity averaged annihilation cross-section ($\langle\sigma v\rangle$) and the energy spectrum ($dN_\gamma/dE$), and another, involving the DM density $\rho_{DM}$, which depends on the astrophysics, the called J-factor.

Therefore, one of the key ingredients to compute the expected gamma-ray flux is the DM density, whose distribution in the Milky Way is unknown. From dynamical simulations it is possible to build some possible profiles to the DM density, of course with uncertainties \cite{Benito:2020lgu}, however, at the region of the Galactic center, it is even harder to deal with that, since there is a loss of information for $r<3$~kpc, and usually an extrapolation is used \cite{2019JCAP...09..046K}. In general, the most common profiles are the NFW, Burkert and Einasto. In principle, we have freedom to make this choice, and the one used here will be the Einasto profile
\begin{equation}
    \rho_{Ein}(r) = \rho_s \, {\rm exp}\left(-\frac{2}{\alpha}\left[\left(\frac{r}{r_s}\right)^{\alpha}-1\right]\right),
    \label{einasto}
\end{equation}
with $\rho_s$ = 0.079, $r_s = 20$~kpc and $\alpha = 0.17$, which controls the slope of the curve.

\begin{figure}[ht!]
    \centering
    \includegraphics[width=0.48\textwidth]{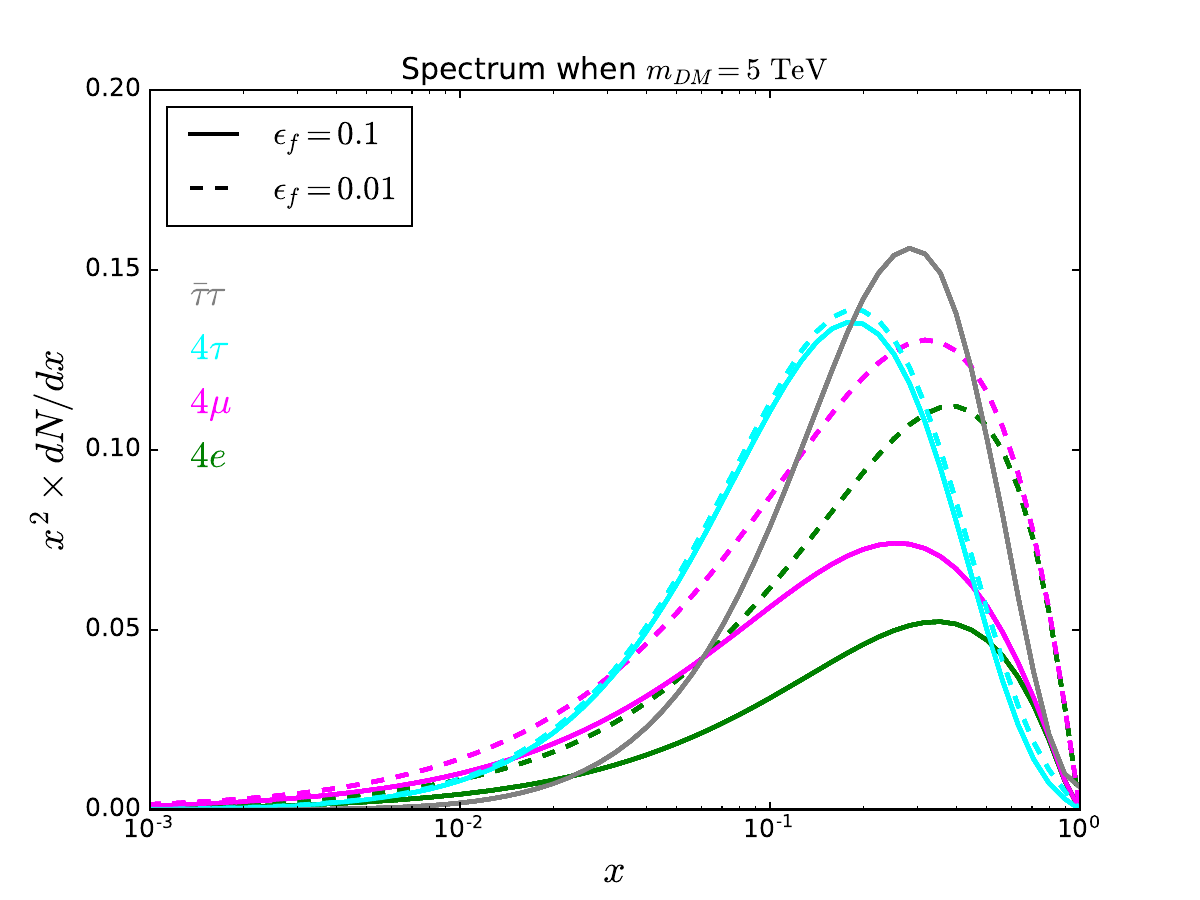}
    \includegraphics[width=0.48\textwidth]{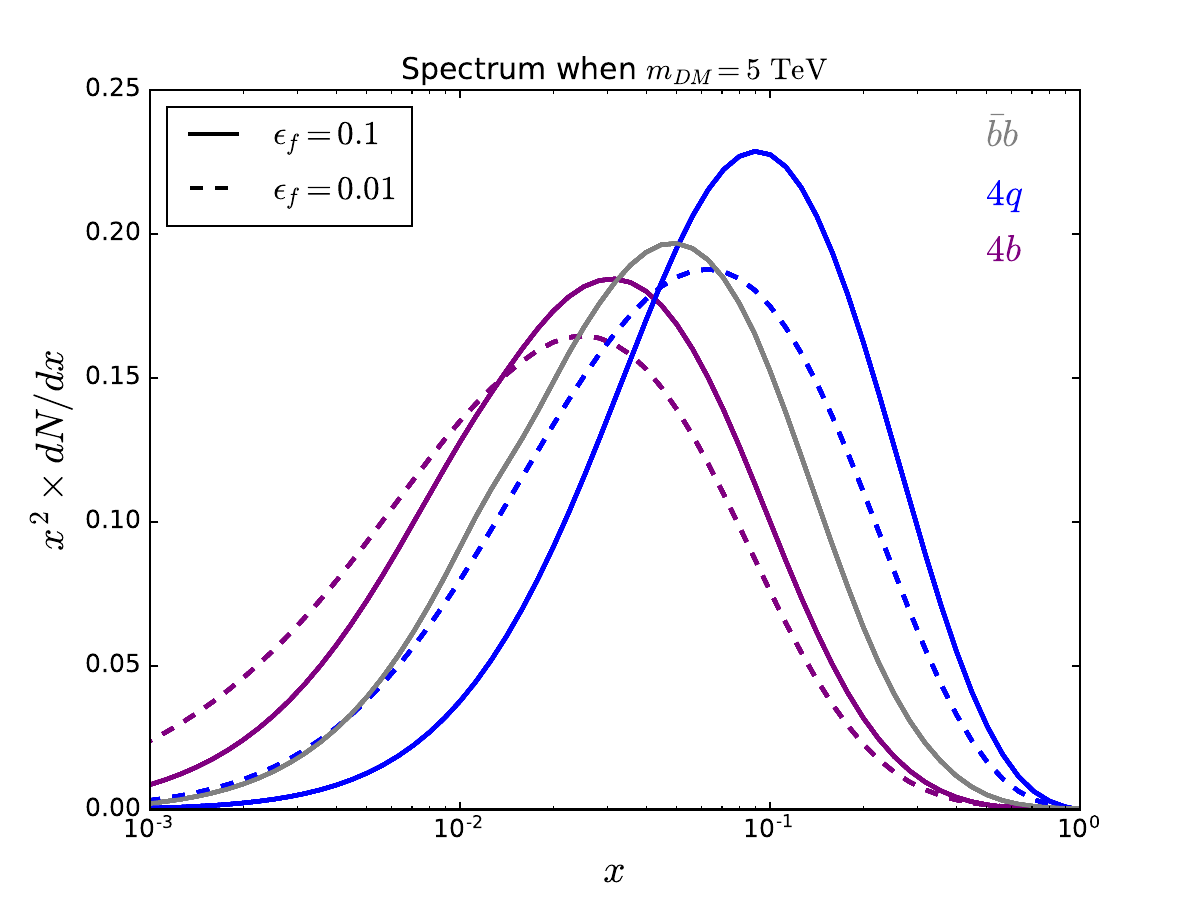}
    \caption{Gamma-ray spectrum for secluded models for leptonic (left) and hadronic channels (left), for a fixed DM mass of $5$~TeV. We choose two values for the $\epsilon$ parameter, $0.1$ (continuous lines) and $0.01$ (dashed lines). For comparison, we include the direct $\Bar{\tau}\tau$ channel in the left plot and the $\Bar{b}b$ channel in the right one.}
    \label{fig:spec}
\end{figure}

The other ingredient is the gamma-ray spectrum ($dN_\gamma/dE$), which depends on the annihilation channel, for our analysis  we choose the DM annihilating into two light metastable mediators which decays into SM particles. One way to obtain this spectrum is to compute the direct annihilation spectrum using the numerical package PPPC4DMID \cite{Cirelli:2010xx}, at the rest frame of the mediator mass $m_V$, the $x_0=2 E_0/m_V$, and then make a boot to the mass of the DM particle mass:  
\begin{equation}
    \frac{dN^\gamma}{dx_1} = 2 \int^{t_{1,max}}_{t_{1,min}} \frac{d x_0}{x_0 \sqrt{1-\epsilon_1^2}} \frac{dN^\gamma}{dx_0} 
\end{equation}
with $\epsilon_1=m_V/m_{DM}$, and
\begin{eqnarray}
  t_{1,min} &=& \frac{2 x_1}{E_1^2} \left(1-\sqrt{1-\epsilon_1^2}\right)  \\
  t_{1,max} &=& Min\left[1,\frac{2 x_1}{E_1^2} \left(1+\sqrt{1-\epsilon_1^2}\right)\right] 
\end{eqnarray}
and $E_1$ the final photon energy and $x_1=E_1/m_{DM}$.
We can also define,
\begin{equation}
    \epsilon_f = \frac{2 m_f}{m_V}.
\end{equation}
which is a parameter that relates the mass of the annihilation channel particle and the mass of the mediator. For our purposes, we will take the following annihilation channels: $V\rightarrow2e$, $V\rightarrow2\mu$, $V\rightarrow2\tau$ $V\rightarrow2b$, and $V\rightarrow2q$, with $q$ representing light quarks. We also choose two different values for the $\epsilon$ parameter, $0.1$ and $0.01$, for each annihilation channel. It is important to stress that we compute the spectrum for scalar mediators (as shown in Fig.~\ref{fig:spec}, for a DM mass of $5$~TeV), but our results will work also for vector mediators as showed by \cite{Elor:2015tva}.
With these ingredients at hand, we will follow to discuss the different experiments used in this analysis.

\section{Gamma-ray observatories}
\label{sec:exp}

Currently, the main gamma-ray experiments which provide the strongest limits on the DM scenarios are the Fermi-LAT satellite, the NASA gamma-ray telescope, and the H.E.S.S. telescope, they probe different energy ranges, what makes them complementary. The next generation of gamma-ray telescopes are the \textit{Cherenkov Telescope Array}, and the \textit{Southern wide field-of-view gamma-ray Observatory} (SWGO), both are expected to probe high energies, at the TeV scale. They will be described in more detail in the following:

\begin{itemize}
    \item Fermi-LAT: able to probe an energy range from $500$~MeV to $500$~GeV, the Fermi-LAT satellite brought a new era in the search for DM particles. The experiment was able to probe the thermal relic cross-section for DM masses lighter than $100$~GeV (for the $\bar{b}b$ channel), looking at dwarf spheroidal galaxies. For this analysis, we will use a sample of 15 dwarfs, and seven years of observation time \cite{Profumo:2017obk}.
    \item H.E.S.S.: it is a ground-based \textit{imaging atmospheric Cherenkov telescope} (IACT) arrays, located in the Khomas Highland of Namibia \cite{Abdallah:2016ygi}, it has been monitoring the Galactic center for the last sixteen years. Able to probe an energy interval of 230~GeV to 30~TeV, the most stringent limits on the annihilation cross-section of TeV WIMPs come from this experiment. For this work, we will use 250h live time looking at the inner $1^\circ$ region of the Galactic halo. It is important to say that we exclude the central region of the galaxy with $|b|<0.3^\circ$, to avoid the strong background that comes from this central region.   
    \item CTA: currently under construction, it is also an IACT, the CTA will be situated on both hemispheres, the Northern in La Palma, Canary Islands, Spain and the Southern in Paranal, Chile. It is expected to probe an energy interval from $20$~GeV to $300$~TeV \cite{Acharyya:2020sbj}, and it will be able, for the first time, to probe the region of the thermal averaged cross-section for DM masses at TeV-scales. In this study, we will use 500~hours of observation time, and also consider a circular $1^\circ$ region around the GC, excluding also the Galactic latitude $\pm 0.3^\circ$.  
    \item SWGO: in the research and development ($R\&D$) phase, it is planned to be located at the South America. SWGO is a water Cherenkov particle detector array, with a wide field of view, and high duty cycle. It will have an energy sensitivity of $500$~GeV to $2$~PeV. It is expected to have an unprecedented sensitivity to multi-TeV DM masses. In this work, we will use ten years of observation time, looking at $10^\circ$ of the central region of the Galactic center, again excluding the central region $|b|<0.3^\circ$.   
\end{itemize}

\section{Results}
\label{sec:res}

To compute the limits/sensitivity on the velocity averaged cross-section ($\langle \sigma v \rangle$) versus the DM mass, we  have to compare the signal versus background. To make it, we follow previous studies and used an ON-OFF 2D joint-likelihood method, in energy and space. 
\begin{figure}[ht!]
    \centering
    \includegraphics[width=0.46\textwidth]{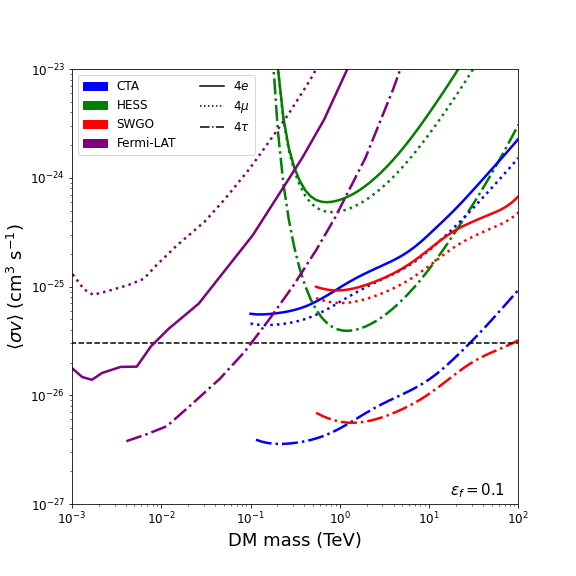}
    \includegraphics[width=0.46\textwidth]{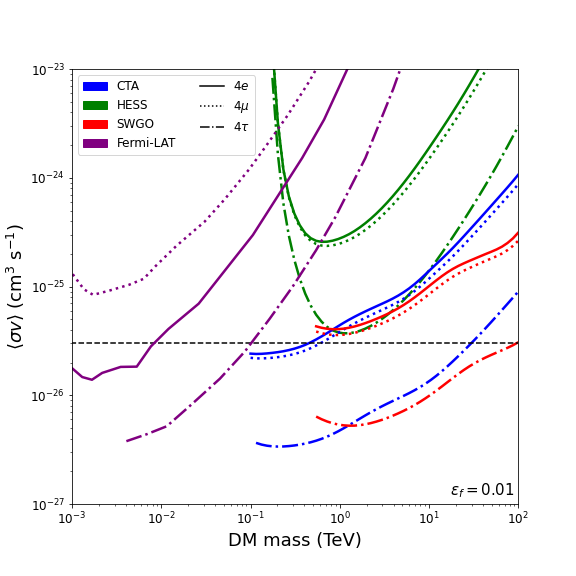}\\
    \includegraphics[width=0.46\textwidth]{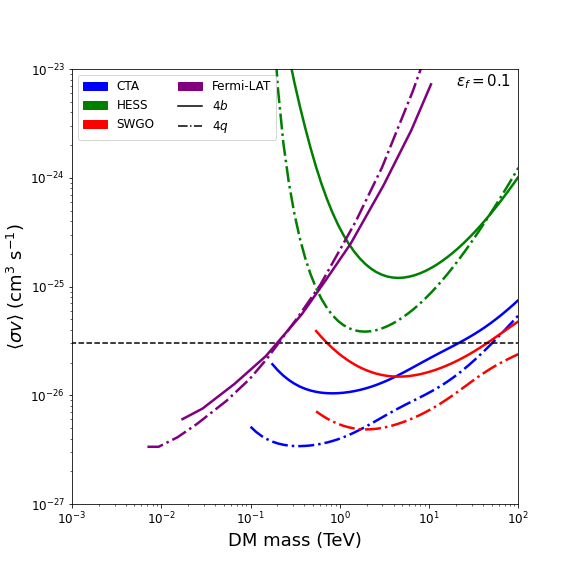}
    \includegraphics[width=0.46\textwidth]{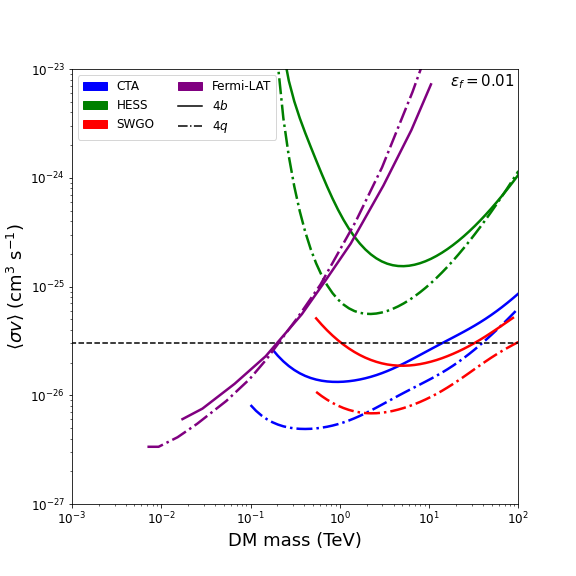}
    \caption{Limits/sensitivity on the velocity averaged cross-section \textit{versus} DM mass. Leptonic channels (top), hadronic channels (bottom)and the values for the $\epsilon_f$ parameter are 0.1 (left) and 0.01 (right). We overlap the results for Fermi-LAT (purple), H.E.S.S. (green), CTA (blue), and SWGO (red). the black dashed line represents the thermal averaged cross-section which provides the right relic abundance. }
    \label{fig:results}
\end{figure}

In Fig.~\ref{fig:results}, we present our results for secluded DM models, combining four different experiments, Fermi-LAT (purple) \cite{Profumo:2017obk}, H.E.S.S. (green), CTA (blue), and SWGO (red). Covering a wide range of DM masses, from $10^{-3}$ to $10^2$~TeV and for $\epsilon_f=0.1$ (left) and $\epsilon_f=0.01$ (right). The leptonic channels (top) leads to less stringent limits, due to be dominated by final state radiation (FSR), except for the $\bar{\tau}\tau$ channel which quickly hadronize producing pions. In general, we observe that for $\bar{e}e$ and $\bar{\tau}\tau$ channels it is possible to reach the thermal annihilation cross-section (which provides the right relic density), given by the black dashed lines in all plots. We can also see that the impact of the change in the mediator's mass is mild. For the hadronic channels (bottom), due to the quick hadronization producing neutral pions, which quickly decay into gamma-rays, the limits are stronger. It is important to mention that the future CTA and SWGO experiments will be able to probe a region at TeV-scales that was never reached before. The improvement in sensitivity reaches around one order of magnitude.

\section{Conclusions}

In this talk, we explored the current and future limits on secluded DM models. For the analyses, we used a 2D joint likelihood method to derive either the limits or the sensitivity of four experiments, namely Fermi-LAT and H.E.S.S. (current), CTA and SWGO (projections). We explored several possible annihilation channels, including  $V\rightarrow2e$, $V\rightarrow2\mu$, $V\rightarrow2\tau$ $V\rightarrow2b$, and $V\rightarrow2q$, and we chose the parameter $\epsilon= 2m_f/m_V$ to be $\epsilon_f=0.1$ and $\epsilon_f=0.01$, to see the impact of the mediator mass in our results. We showed that the future observatories will be able to probe a large region of the TeV DM masses, reaching almost one order of magnitude in improvement over previous probes.

\bibliographystyle{unsrt}
\bibliography{ref.bib}

\end{document}